D. A. Kulikov, R. S. Tutik[*]

*Dniepropetrovsk National University, Department of Physics,
13 Naukova str., Dnieprorpetrovsk, 49050 Ukraine*

# RENORMALIZATION OF EXPANSIONS FOR REGGE TRAJECTORIES OF THE SCHRÖDINGER EQUATION

*Abstract* — A recursion technique for the renormalization of semiclassical expansions for the Regge trajectories of bound states of the Schrödinger equation is developed. As an application of the proposed technique, the two-parameter renormalization scheme for the Regge trajectories of the bound states in the Martin potential is considered.

## 1. INTRODUCTION

An important problem in non-relativistic quantum mechanics is the study of bound states of the radial Schrödinger equation

$$-\frac{\hbar^2}{2m}u''(r) + [\frac{\hbar^2 l(l+1)}{2mr^2} + V(r)]u(r) = Eu(r). \qquad (1)$$

For this equation, numerous methods have been developed to calculate the eigenenergies $E_{nl}$ with integer orbital ($l$) and radial ($n$) quantum numbers. Sometimes, for example within the framework of the potential models of hadrons, it is, however, more convenient to reformulate the problem and to calculate the dependence $l$ on $E$, i. e. the main ($n=0$) and daughter ($n=1,2,...$) Regge trajectories

$$\alpha(E) = \hbar l(E). \qquad (2)$$

Some years ago a powerful semiclassical technique for deriving the Regge trajectories of the bound-state problem of the Schrödinger equation was proposed [1]. This technique enables to obtain the analytic expressions for the Regge trajectories as the explicit series expansions in powers of Planck's constant.

But the obtained expansions are the asymptotical ones and, as it is shown by calculations, give sufficiently accurate results in the case of the only parent Regge trajectory ($n=0$). The goal of the present work is to develop the renormalization techinique for improving the convergence of the expansions for daughter Regge trajectories.

## 2. RECURRENT RELATIONS FOR REGGE TRAJECTORIES

Let us consider the radial Schrödinger equation (1) with the boundary conditions $u(0) = u(\infty) = 0$. Following the work [1], with the use of the substitution for the logarithmic derivative, $C(r) = \hbar u'(r)/u(r)$, we transform Eq.(1) into the non-linear Riccati equation

$$\hbar C'(r) + C^2(r) = 2m(E - V(r)) + \frac{1}{r^2}\alpha(E)(\alpha(E) + \hbar). \qquad (3)$$

---

[*] E-mail address: tutik@ff.dsu.dp.ua

A solution to this equation and the corresponding Regge trajectories will be sought in the form of asymptotical expansions in powers of Planck's constant

$$C(r) = \sum_{k=0}^{\infty} C_k(r)\hbar^k, \qquad \alpha(E) = \sum_{k=0}^{\infty} \alpha_k(E)\hbar^k. \qquad (4)$$

The basic point of the method is the passage to the classical limit by the rule

$$\hbar \to 0, \ n = const, \ l \to \infty, \ \hbar n \to 0, \ \hbar l = const. \qquad (5)$$

As a consequence, in the classical limit the particle is located at the bottom of the potential well and thus moves along the stable circular orbit of the radius $r_0$. Then, the zeroth approximation to the quantum-mechanical Regge trajectory is given by the classical angular momentum of the particle $\alpha_0(E) = (mr_0^3 V'(r_0))^{1/2}$. To calculate the quantum corrections, we expand the functions $C_k(r)$ and $V(r)$ in the Laurent series in powers of the deviation from the minimum of effective potential, $x = (r - r_0)/r_0$,

$$C_k(r) = x^{1-2k} \sum_{i=0}^{\infty} C_i^k x^i, \qquad V(r) = \sum_{i=0}^{\infty} V_i x^i. \qquad (6)$$

After this all calculations are reduced to algebraic operations.

The application of the argument principle, known from the complex analysis, permits us to take into account the nodes of the wave function as

$$\frac{1}{2\pi i} \oint C(r)dr = n\hbar, \quad n = 0,1,2,\ldots \qquad (7)$$

where $n$ is a number of the zeros of the wave function on the real axis, and the integration contour encloses only the above nodes.

Eq.(5) enables to rewrite the last condition through the coefficients $C_i^k$

$$C_0^1 = n/r_0, \quad C_{2k-2}^k = 0, \ k \neq 1. \qquad (8)$$

In result, after substituting the series (4) and (6) in Eq.(3) and collecting the coefficients at the same orders, we have the recurrent relation [1] for computing the coefficients of the expansions (4) for Regge trajectories

$$\alpha_k(E) = \frac{1}{2\alpha_0}[\alpha_{k-1} + \sum_{j=1}^{k-1}\alpha_j\alpha_{k-j} - r_0 C_{2k-2}^{k-1} - r_0^2 \sum_{j=0}^{k}\sum_{p=0}^{2k-2} C_p^j C_{2k-2-p}^{k-j}] \qquad (9)$$

where the coefficients $C_i^k$ of the Laurent series (6) obey the relations:

$$C_0^0 = -\omega, \quad C_i^0 = \frac{1}{2\omega}(\sum_{j=1}^{i-1} C_j^0 C_{i-j}^0 - \omega^2 a_i), \ i \geq 1, \qquad (10)$$

$$C_i^k = \frac{1}{2C_0^0}\{-\frac{i-2k+3}{r_0}C_i^{k-1} - \sum_{j=1}^{k-1}\sum_{p=0}^{i} C_p^j C_{i-p}^{k-j} - 2\sum_{p=1}^{i} C_p^0 C_{i-p}^k +$$

$$+ \theta(i-2k+2)\frac{(-1)^i(i-2k+3)}{r_0^2}(\alpha_{k-1} + \sum_{j=0}^{k}\alpha_j\alpha_{k-j})\}, \ k > 0, \ i \neq 2k-2, \qquad (11)$$

and $\omega$ and $a_i$ are expressed through the parameters of the potential as

$$\omega^2 = 2m(V_2 + \frac{3}{2}V_1), \quad a_i = \frac{2m}{\omega^2}(V_{i+2} + (-1)^i \frac{3+i}{2}V_1). \tag{12}$$

The obtained recurrent formulae completely resolve the problem of deriving the parent and daughter Regge trajectories for the bound states of the radial Schrödinger equation.

### 3. RENORMALIZATION OF EXPANSIONS FOR REGGE TRAJECTORIES

The expansions for Regge trajectories, obtained in previous Section, are the asymptotical ones and in the general case these expansions diverge. For improving the approximation, the renormalization technique can be used [2-4]. The renormalization of the expansion is based on a redistribution of series items, with adding artificial parameters (renormalization parameters), that leads to the partial summation of the series and thus to the improvement of the convergence.

Let us consider the mass renormalization. For this, in the initial equation (2) we replace the mass $m$ by its expansion in powers of $\hbar$

$$m = \sum_{k=0}^{\infty} m_i \hbar^i \tag{13}$$

that involves the renormalization parameters $m_i$ in expansion for the Regge trajectory

$$\tilde{\alpha}^{(N)}(E, m_i) = \sum_{k=0}^{N} \tilde{\alpha}_k(E, m_i) \hbar^k. \tag{14}$$

Coefficients $\tilde{\alpha}_k(E, m_i)$ are calculated through the following recurrent relations which are the modifications of Eqs.(9) and (11)

$$\tilde{\alpha}_k(E, m_i) = \frac{1}{2\tilde{\alpha}_0}[\tilde{\alpha}_{k-1} + \sum_{j=1}^{k-1}\tilde{\alpha}_j\tilde{\alpha}_{k-j} - r_0 C_{2k-2}^{k-1} -$$
$$- r_0^2 \sum_{j=0}^{k}\sum_{p=0}^{2k-2} C_p^j C_{2k-2-p}^{k-j} - m_k r_0 V_1 ], \tag{15}$$

$$C_i^k = \frac{1}{2C_0^0}\{-\frac{i-2k+3}{r_0}C_i^{k-1} - \sum_{j=1}^{k-1}\sum_{p=0}^{i}C_p^j C_{i-p}^{k-j} - 2\sum_{p=1}^{i}C_p^0 C_{i-p}^k$$
$$+ \theta(i-2k+2)[\frac{(-1)^i(i-2k+3)}{r_0^2}(\tilde{\alpha}_{k-1} + \sum_{j=0}^{k}\tilde{\alpha}_j\tilde{\alpha}_{k-j}) + 2m_k V_{i-2k+2}]\} \tag{16}$$

whereas Eq.(10) remains unchanged, and in Eq.(12) $m$ must be replaced by $m_0$.

From the mathematical point of view, such renormalization corresponds to the transition from the asymptotical expansion according to Poincaré to the asymptotical expansion according to Erdélyi [5]. Its physical meaning consists in the transition to another zeroth approximation, i. e. another classical angular momentum of the particle. This change of the classical angular momentum is permitted because in quantum mechanics the rule of passage to the classical limit is not unambiguously defined.

Now let us discuss ways of finding the optimal values of the renormalization parameters. Usually the «principle of minimal sencitivity» [2] is used. A reason for this principle is that the physical quantities should not depend on artificial parameters. Hence, the derivatives with respect to these parameters must equal to zero. An alternative condition, the «principle of the fastest convergence», consists in equaling to zero the last calculated series items. Note that the obtained values of the renormalization parameters depend on the order of approximation.

As an illustration of the method, we consider the proposed renormalization for Regge trajectories in the case of the Schrödinger equation with the Martin potential, $V(r) = Ar^{0,1}$, widely used in the hadron spectroscopy.

For the power-law potential, $V(r) = Ar^\nu$, the Regge trajectories, with the corrections up to order $\hbar^4$, have the form

$$\alpha^{(4)}(E) = \alpha_0(E) - \frac{\hbar}{2}(1 + q\sqrt{\nu+2}) + \frac{\hbar^2}{288\alpha_0(E)}(\nu-2)(\nu+1)(3q^2-1) -$$
$$- \frac{\hbar^3}{13824\alpha_0^2(E)\sqrt{\nu+2}}(\nu-2)(\nu+1)[(5\nu^2 - 29\nu - 58)q^3 - (\nu^2 - 25\nu - 50)q] +$$
$$+ \frac{\hbar^4}{2985980(\nu+2)\alpha_0^3(E)}(\nu-2)(\nu+1)[(2415\nu^4 - 70170\nu^3 + 24615\nu^2 + \quad (17)$$
$$+ 659820\nu + 659820)q^4 + (3270\nu^4 + 59340\nu^3 - 138330\nu^2 - 1028040\nu -$$
$$- 1028040)q^2 - 613\nu^4 + 974\nu^3 + 46947\nu^2 + 179996\nu + 179996]$$

where $q = 2n+1$, $\alpha_0(E) = \sqrt{\nu m E}(2E/A(\nu+2))^{(\nu+2)/2\nu}$.

In this work we study the two-parameter renormalization schemes and truncate the expansion (13) up to the equality $m = m_0 + m_1\hbar + m_2\hbar^2$. The values of independent parameters $m_1$ and $m_2$ are fixed according to the «principle of minimal sencitivity»

$$\partial \tilde{\alpha}^{(4)} / \partial m_1 = 0, \quad \partial \tilde{\alpha}^{(4)} / \partial m_2 = 0 \quad \text{(scheme 1)} \qquad (18)$$

and according to the «principle of the fastest convergence»

$$\tilde{\alpha}_3 = 0, \qquad \tilde{\alpha}_4 = 0 \qquad \text{(scheme 2)}. \qquad (19)$$

The results of our calculations (in units with $\hbar = c = 1$, $A = 1$) for the fourth order of approximation are presented in Table 1.

**Table 1.** Regge trajectories for the Schrödinger equation with the Martin potential. The values of trajectories are calculated at the energies which correspond to the bound states with integer $n$ and $l = \alpha_{exact}$.

| $n$ | $\alpha_{exact}$ | $\alpha^{(4)}$ unrenormalized | $\tilde{\alpha}^{(4)}$ scheme 1 | $\tilde{\alpha}^{(4)}$ scheme 2 |
|---|---|---|---|---|
| 1 | 0 | 0,00622 | 0,00250 | 0,00165 |
| 1 | 1 | 1,00152 | 1,00051 | 1,00032 |
| 2 | 0 | 0,02022 | 0,01001 | 0,00726 |
| 2 | 1 | 1,00724 | 1,00301 | 1,00205 |
| 3 | 0 | 0,03699 | 0,01977 | 0,01487 |
| 4 | 0 | 0,05512 | 0,03908 | 0,02357 |

From Table 1 we see that the values of Regge trajectories obtained through the renormalized expansions (two last columns in the table) are more accurate than the unrenormalized ones (the third column). Notice that the both schemes provide practically the same accuracy.

### 4. CONCLUSION

In the work a recursion technique for the renormalization of the semiclassical expansions for Regge trajectories of the Schrödinger equation has been developed. Inclusion of artificial parameters directly in the recurrent formulae enables to use not only one-parameter renormalization schemes but also many-parameter ones. As an example of application, we have considered the two-parameter renormalization of the Regge trajectories for bound states of the Schrödinger equation with the Martin potential. It has been shown that the renormalized expansions provide a more accurate approximation in the case of daughter Regge trajectories. The results of this work can be extended to the recursion schemes of deriving the Regge trajectories for bound states of the relativistic Klein-Gordon and Dirac equations.